\documentclass[twocolumn,superscriptaddress]{revtex4}
\usepackage{graphicx}
\usepackage{times}
\usepackage{amsmath}
\usepackage{amssymb}
\usepackage{units}
\usepackage{stmaryrd} 
\usepackage[compact]{titlesec}
\titlespacing{\section}{0pt}{*0}{*0}
\titlespacing{\subsection}{0pt}{*0}{*0}
\titlespacing{\subsubsection}{0pt}{*0}{*0}

\begin{document}
\preprint{0}

\title{ARPES insights on the metallic states of YbB$_{6}$(001):\\ E(k) dispersion, temporal changes and spatial variation}

\author{E. Frantzeskakis}
\email{e.frantzeskakis@uva.nl} 
\address{Institute of Physics (IoP), University of Amsterdam, Science Park 904, 1098 XH, Amsterdam, the Netherlands}

\author{N. de Jong}
\address{Institute of Physics (IoP), University of Amsterdam, Science Park 904, 1098 XH, Amsterdam, the Netherlands}

\author{J. X. Zhang}
\address{Dept. of Materials Science and Engineering, Beijing University of Technology, Pingleyuan100, Chaoyang Districts, Beijing 100124, China}
\address{Dept. of Materials Science and Engineering, Hefei University of Technology, Tunxi road 193, Anhui, Hefei 230009, China}

\author{X. Zhang}
\address{Dept. of Materials Science and Engineering, Beijing University of Technology, Pingleyuan100, Chaoyang Districts, Beijing 100124, China}

\author{Z. Li}
\address{Dept. of Materials Science and Engineering, Hefei University of Technology, Tunxi road 193, Anhui, Hefei 230009, China}

\author{C. L. Liang}
\address{Dept. of Materials Science and Engineering, Beijing University of Technology, Pingleyuan100, Chaoyang Districts, Beijing 100124, China}

\author{Y. Wang}
\address{Dept. of Materials Science and Engineering, Beijing University of Technology, Pingleyuan100, Chaoyang Districts, Beijing 100124, China}

\author{A. Varykhalov}
\address{Helmholtz-Zentrum Berlin f\"{u}r Materialien und Energie, Albert-Einstein-Strasse 15, 12489 Berlin, Germany}

\author{Y. K. Huang}
\address{Institute of Physics (IoP), University of Amsterdam, Science Park 904, 1098 XH, Amsterdam, the Netherlands}

\author{M. S. Golden}
\email{m.s.golden@uva.nl} 
\address{Institute of Physics (IoP), University of Amsterdam, Science Park 904, 1098 XH, Amsterdam, the Netherlands}


\begin{abstract}
We report high resolution Angle Resolved PhotoElectron Spectroscopy (ARPES) results on the (001) cleavage surface of YbB$_{6}$, a rare-earth compound which has been recently predicted to host surface electronic states with topological character.
We observe two types of well-resolved metallic states, whose Fermi contours encircle the time-reversal invariant momenta of
the YbB$_{6}$(001) surface Brillouin zone, and whose full (E,$k$)-dispersion relation can be measured wholly unmasked by states from the rest of the electronic structure. 
Although the two-dimensional character of these metallic states is confirmed by their lack of out-of-plane dispersion, two new aspects are revealed in these experiments.
Firstly, these states do not resemble two branches of opposite, linear velocity that cross at a Dirac point, but rather straightforward parabolas which terminate to high binding energy with a clear band bottom.
Secondly, these states are sensitive to time-dependent changes of the YbB$_{6}$ surface under ultrahigh vacuum conditions.
Adding the fact that these data from cleaved YbB$_{6}$ surfaces also display spatial variations in the electronic structure, it appears there is little in common between the theoretical expectations for an idealized YbB$_{6}$(001) crystal truncation on the one hand, and these ARPES data from real cleavage surfaces on the other.

\end{abstract}

\maketitle

\section*{Introduction}

Soon after the experimental discovery of topological insulators (TIs) -materials where surface conduction electrons possess unconventional characteristics related to the topology of the underlying bulk electronic structure \cite{Hasan2010}- a strong scientific interest has been focused on combining their unique electronic properties with strong electron correlations.
Apart from the realization of new exotic physical phenomena such as the topological Mott phase \cite{Pesin2010} or the fractional TIs \cite{Sheng2011, Regnault2011}, the combined effect of strong electron correlations and topological characteristics may give rise to novel materials with a high potential for emerging applications.
Prime examples are the topological superconductors \cite{Qi2011} and the idea of topological Kondo insulators (TKIs) \cite{Dzero2010, Alexandrov2013, Lu2013}.

SmB$_{6}$ is considered as the first candidate material for a TKI. Transport \cite{Wolgast2013, Li2013, Fisk2013, Zhang2013, Fisk2014}, Angle-Resolved Photoemission (ARPES) \cite{Hasan2013, Ming2013, Feng2013, Frantzeskakis2013, Damascelli2013, Denlinger2013, Reinert2014} and Scanning Tunneling Microscopy/Spectroscopy (STM/STS) \cite{Hoffman2013, Rossler2014, Ruan2014}
studies have intensively sought for clear-cut signatures of topological characteristics in SmB$_{6}$.
The ARPES data from the different studies resemble each other but the interpretations vary: the majority are in favor \cite{Hasan2013, Ming2013, Feng2013, Denlinger2013, Reinert2014}, but some are more sceptical \cite{Frantzeskakis2013, Damascelli2013} as to whether the observed bands at the Fermi surface of cleaved crystals are the topological surface states whose theoretical prediction sparked such a huge experimental effort. 
Surface electronic structure studies on the nanoscale using STS have - as yet - been unable to uncover any quasiparticle interference signals from surface states in samarium hexaboride (unlike the situation in the Bi-based 3D TI systems \cite{Roushan2009,Zhang2009,Alpichshev2010,Beidenkopf2011}), and topographic studies have shown there to be a variety of terminations on the surface of UHV-cleaved SmB$_{6}$ crystals \cite{Hoffman2013, Rossler2014, Ruan2014}.        

Given the debate on the samarium compound, a recent theoretical study on YbB$_{6}$ has garnered much interest \cite{Dai2014}. It was proposed that the (001) and (111) cleavage surfaces of YbB$_{6}$ could also host electronic states of topological character provided that YbB$_{6}$ is mixed valent.
In these calculations, the energy gap in which the topological surface states should be situated is proposed to be 31 meV, some three times larger than similar calculations predict for its samarium analogue \cite{Lu2013}.
     \begin{figure*}
  \centering
  \includegraphics[width = 14.5 cm]{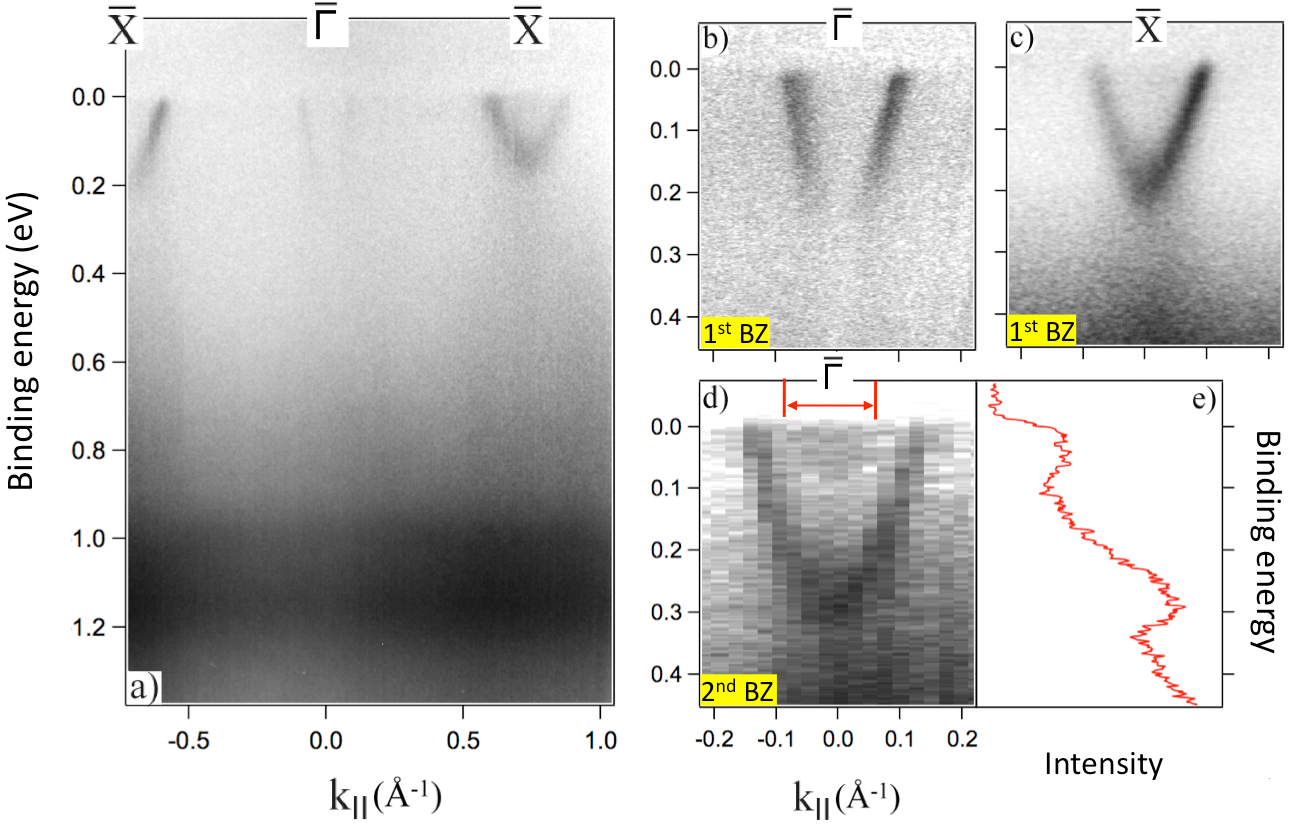}
  \caption{
(a) Electronic band structure of YbB$_{6}$(001) along the $\overline{\textmd{X}\Gamma\textmd{X}}$ direction of the surface Brillouin zone. The non-dispersive feature at $\sim$1.1 eV corresponds to the lowest binding energy Yb 4$f$ states. Electron pockets are observed at $\overline{\Gamma}$ and $\overline{\textmd{X}}$. (b) Zoom of the electronic states near E$_{\textmd{F}}$ at $\overline{\Gamma}$. (c) Zoom of the electronic states near E$_{\textmd{F}}$ at $\overline{\textmd{X}}$. (d)
Same as panel (b), but acquired in the 2$^{\textmd{nd}}$ SBZ, where favorable photoemission matrix elements give clear access to the band minimum. A second electron pocket is also observed. (e) Energy distribution curve for $k$ around $\overline{\Gamma}$, taken by integrating the area denoted by the red arrow in panel (d), which is well inside the $k_{\textmd{F}}$ values for the larger electron pocket. All data in this figure has been acquired using an excitation energy of 35 eV at a sample temperature of 37K.}
\label{Fig1}
\end{figure*}

YbB$_{6}$ crystallizes in the same CsCl-type crystal structure as SmB$_{6}$ \cite{Aprea2013, Tarascon1980}. 
In contrast to the mixed-valent character for the bulk Sm ions in SmB$_{6}$, bulk sensitive susceptibility measurements suggest a purely divalent
character for YbB$_{6}$ \cite{Tarascon1980,Nanba1993}. This is in good agreement with optical spectroscopy data where only 1\% of the Yb ions are in the
trivalent state
\cite{Nanba1993}, thus classifying YbB$_{6}$ as a divalent hexaboride. The materials family of the divalent hexaborides was investigated intensively after the
discovery of ferromagnetism in La-doped CaB$_{6}$ in 1999 \cite{Young1999}. The issue as to whether hole and electron like bands cross near the X-point
in $k$-space so as to enable formation of an excitonic insulator \cite{Zhitomirsky1999,Balents2000,Barzykin2000,Murakami2002} was an important part of
the debate at that time. From a combination of ARPES and bulk-sensitive, $k$-resolved, resonant inelastic x-ray scattering, a gap between the valence and
conduction bands at the X-point of more than 1 eV was concluded for the divalent hexaborides of Ca, Sr and the rare earth Eu \cite{Denlinger2002_2}. Going
back to the claims for a mixed-valent -rather than divalent- character of YbB$_{6}$ \cite{Dai2014}, they were based on the results of x-ray electron
spectroscopies which found the Yb valence to be 2.2 \cite{Nanba1993}. However, seeing as the trivalent Yb component is only clearly detected in surface-sensitive electron spectroscopies \cite{Nanba1993,Kakizaki1993}, it could well come from a surface-related electronic structure that differs from that of the divalent bulk.
Since mixed-valent character for the bulk of YbB$_{6}$ cannot be supported experimentally, the general applicability of theoretical calculations treating YbB$_{6}$
as mixed valent in the bulk \cite{Dai2014} may need further clarification. In addition, we mention that recent magnetoresistance measurements failed to identify either a mixed
valence state or a surface conduction channel in the temperature-dependent resistivity of YbB$_{6}$ \cite{Fisk2014}. This is in stark contrast with SmB$_{6}$, in which both bulk mixed valent character \cite{Curnoe2000,Mizumaki2009} and a surface contribution to resistivity \cite{Zhang2013,Fisk2013,Wolgast2013} have been well established. 
\begin{figure*}
  \centering
  \includegraphics[width =11.7 cm]{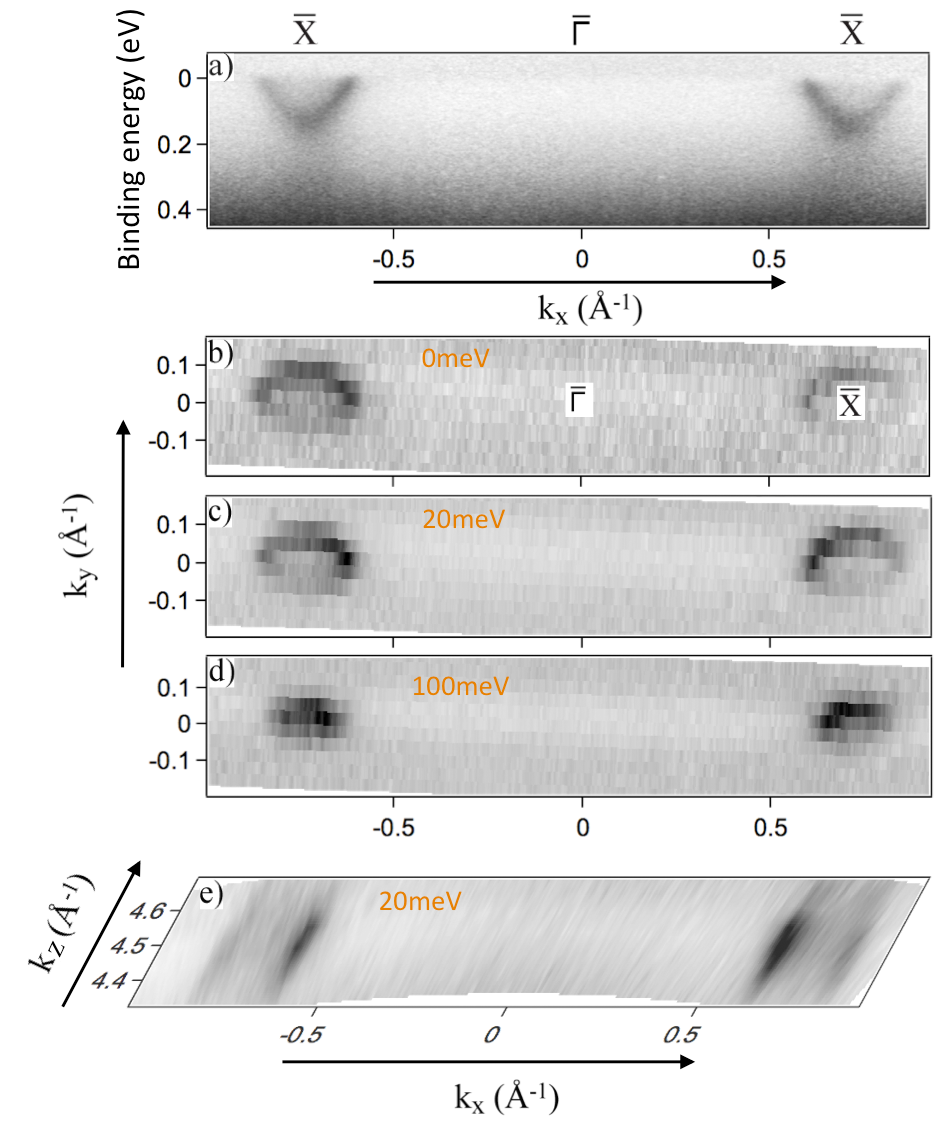}
  \caption{
(a) E($k$) dispersion of the states at $\overline{\textmd{X}}$ along $\overline{\textmd{X}\Gamma\textmd{X}}$ acquired with $h\nu=70$ eV. (b)-(d) In-plane constant energy contours: ($k_{\textmd{x}}$, $k_{\textmd{y}}$) momentum distribution of the electron pockets centred at $\overline{\textmd{X}}$ (b) at the Fermi level, (c) at a binding energy of 20 meV and (d) at a binding energy of 100 meV. (e) Out-of-plane constant energy contours: ($k_{\textmd{x}}$, $k_{\textmd{z}}$) momentum distribution of the electron pockets centered at  $\overline{\textmd{X}}$ at a binding energy of 20 meV. Data in panels (a) through (d) has been acquired using an excitation energy of 70 eV. Panel (e) is based upon data recorded with photon energies between 62 and 78 eV. In all cases, the sample temperature was 37K.}
\label{Fig2}
\end{figure*}

Interestingly, very recent ARPES studies - inspired by the theoretical predictions \cite{Dai2014} of topological states in mixed valent YbB$_{6}$ - have identified metallic states on the (001) cleavage surface of YbB$_{6}$ \cite{Feng2014, Ming2014, Hasan2014}.
In agreement with PES data from 20 years earlier \cite{Kakizaki1993}, all three recent studies show the binding energy of the lowest lying occupied Yb 4$f$ state in YbB$_{6}$ to be ca. 1 eV, thus of order 1 eV higher than in the theory \cite{Dai2014}.
This discrepancy has been shown to be reduced when the on-site Coulomb interaction for the Yb 4$f$ states is taken into account \cite{Hasan2014}.
If band inversion near the Fermi energy (E$_{\textmd{F}}$) in the bulk is driving the formation of topologically non-trivial surface states in YbB$_{6}$, then it should occur between Yb 5$d$ band B 2$p$ related states \cite{Hasan2014,Ming2014}, rather than due to 4$f$-5$d$ hybridization \cite{Dai2014}.
Although there is no indication for topological character in divalent YbB$_{6}$ by bulk transport measurements \cite{Fisk2014}, two of the aforementioned ARPES papers argue that the observed metallic states are topological surface states \cite{Feng2014, Ming2014}, while one has some reservations \cite{Hasan2014}.
Our work aims to clarify whether the observed metallic states are of topological nature or not. 

We report high-resolution ARPES results from flux-free, floating zone grown single crystals of YbB$_{6}$ in which metallic, two-dimensional states are detected. 
However, their dispersion relation is not that of two branches of opposite velocity that cross at a Dirac point, but rather that of straightforward parabolas exhibiting clearly resolved band bottoms, unencumbered by intensity from higher lying states.
In addition, the data show that the whole electronic structure is subject to substantial time-dependent changes in which the surface chemical potential shifts downwards by at least 150 meV, and that the electronic properties of the surface vary from different spatial regions of the cleave. 
Taken together, these new experimental insights consist of a non-Dirac-like dispersion (clear band bottom + parabolic dispersion), a substantial temporal evolution and spatial inhomogeneities. As a result, they reveal a complexity in the behavior of the (001) cleavage surface of YbB$_{6}$ that is yet to be captured theoretically, and argue for caution as regards conclusions that ARPES data prove the existence of topological surface states at YbB$_{6}$(001) surfaces.\\ 

\section*{Results and Discussion}

Fig. 1a shows the experimental band structure of YbB$_{6}$(001) as measured by ARPES. The gross features agree with recent
ARPES studies \cite{Feng2014, Ming2014, Hasan2014}, including the fact that the 4$f^{13}$ final state emission from the divalent Yb appears at a binding energy of $\sim$1.1 eV \cite{Kakizaki1993}. 
Shallow features are observed at the $\overline{\Gamma}$ and $\overline{\textmd{X}}$ high-symmetry points of the surface Brillouin zone (SBZ) and they clearly cross the Fermi level, indicating their metallic nature.
Figs. 1b-1d give a closer view of the near-E$_{\textmd{F}}$ electronic structure and reveal that the states at both $\overline{\Gamma}$ and $\overline{\textmd{X}}$ are electron pockets.
The identification of this important characteristic is possible in these data in comparison to the other
available ARPES studies \cite{Feng2014, Ming2014, Hasan2014}, because the residual intensity from higher lying states does not mask these electron pockets at any binding energy.
At first glance, the data of Fig. 1b suggest a linear dispersion, although the crossing point of the two dispersive branches is suppressed. Fig. 1d shows this suppression to be a matrix element effect as in a higher SBZ the state at $\overline{\Gamma}$ reveals itself to be a simple parabola, with a clear band bottom lying -in these data- at 300 meV binding energy.
Fig. 1c shows the state at $\overline{\textmd{X}}$ to have a band bottom at a little below 200 meV in these data.
Taking Figs. 1c ($\overline{\textmd{X}}$) and 1d ($\overline{\Gamma}$) at face value, there would seem to be little or no room for the existence of a Dirac cone in these ARPES data from YbB$_{6}$.
We close the discussion of Fig. 1 by pointing out that k$_{\textmd{F}}$ for the $\overline{\Gamma}$ pocket differs in Figs. 1b and 1d, a point we return to in the context of Fig. 3.
\begin{figure*}
  \centering
  \includegraphics[width = 18 cm]{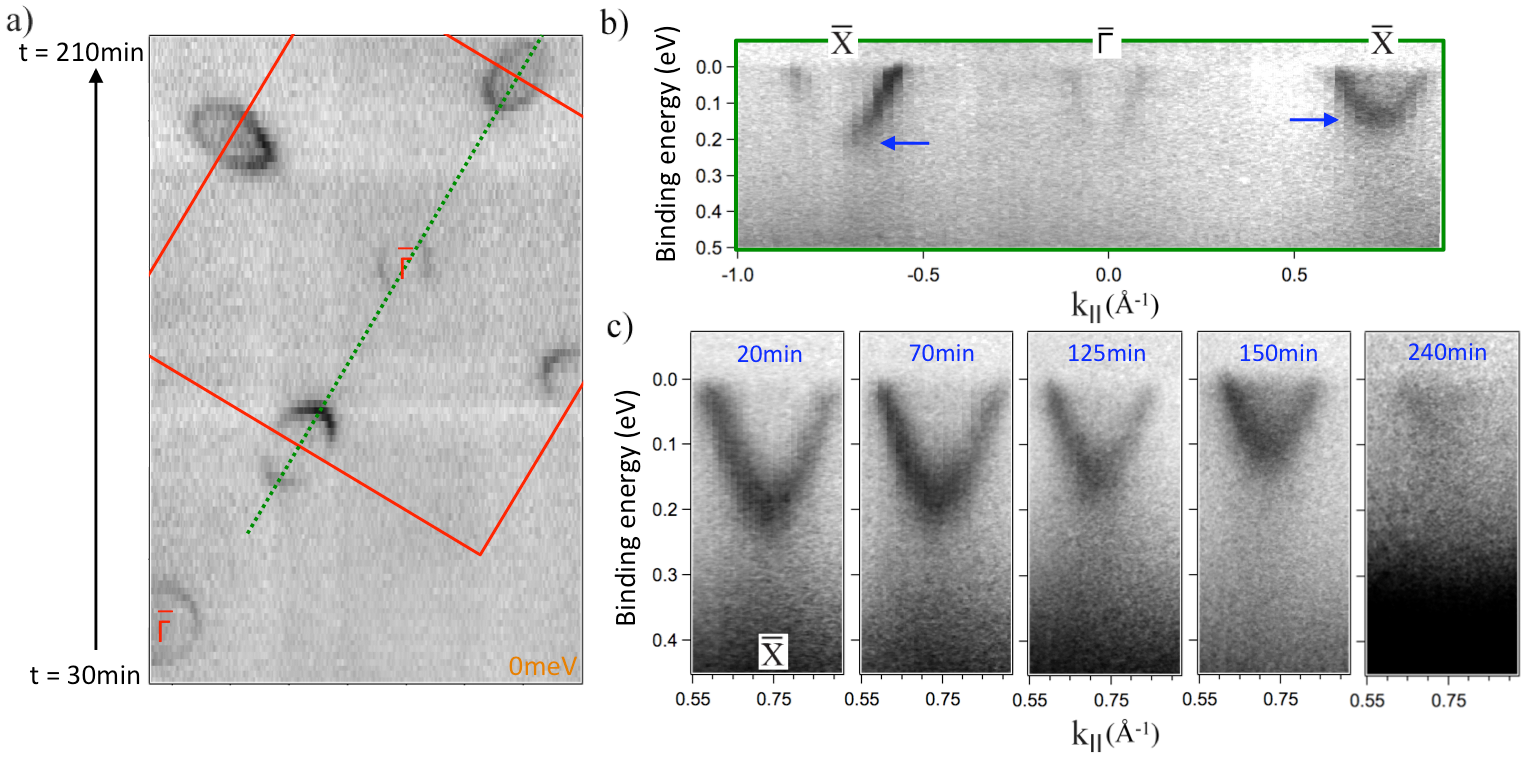}
  \caption{
(a) The Fermi surface of YbB$_{6}$(001). Clear elliptical (circular) contours are seen to close around the $\overline{\Gamma}$ ($\overline{\textmd{X}}$) points.
Data has been acquired by sequential scanning of horizontal $k$-slices using an angular step of 0.5$^{\textmd{o}}$.
The time since sample cleavage is shown on the left for the first and last $k$-slice
The excitation energy was 35 eV and the greyscale shows the ARPES intensity integrated $\pm$5 meV around the Fermi energy.
Note the clear reduction in the size of the constant energy contours during the course of the measurement (from the bottom to the top).
The red, solid lines denote the borders of the 1$^{\textmd{st}}$ surface Brillouin zone.
(b) E($k$) dispersion along the $k$-path indicated in (a) as a green dotted line. Arrows highlight that the electron pocket at positive $k$-values (acquired towards the end of the sequential scanning) has a shallower minimum than that at negative $k$-values (acquired towards the beginning).
(c) Upward shift of the binding energy of the electron pocket centered on $\overline{\textmd{X}}$ as the time since sample cleavage increases.
For all data, the excitation energy was 70 eV and the sample was held at 37K.}
\label{Fig3}
\end{figure*}

We remark that an additional weak electron pocket is observed inside the main $\overline{\Gamma}$ electron pocket (Fig. 1d). The additional electron pocket shows up as a peak close to E$_{\textmd{F}}$ in the energy distribution curve resulting from integration over $k$-values well inside the Fermi surface crossing of the main electron pocket (Fig. 1e).
Such features have been previously observed both at $\overline{\Gamma}$ and $\overline{\textmd{X}}$ points and have been mainly attributed to the conduction
band of YbB$_{6}$(001) \cite{Feng2014, Ming2014, Hasan2014}. In light of the evidence just mentioned against the main electron pockets shown in Figs. 1c and 1d being topological surface states, a different interpretation of this second electron pocket is possible.
It is possible that a fresh cleavage surface of YbB$_{6}$(001) exhibits substantial surface band bending which induces a downward energy shift and -in synergy with the surface potential- creates a potential well. The two electron pockets shown in Fig. 1d could then be the $\nu=1$ and $\nu=2$ quantum well states of the bulk conduction band \cite{Speer2006,Santander2011} which has shifted below E$_{\textmd{F}}$ and is subject to the confinement potential.
In the following, we argue that the ($k_{\textmd{x}}$,$k_{\textmd{y}}$) contours of the metallic states are reminiscent of those measured from the
conduction band of other divalent hexaborides which are known to be subject to downward band bending \cite{Denlinger2002,Denlinger2002_2}, we present experimental evidence that the observed states are indeed confined in the near-surface region
and we will point out a possible origin for the confinement potential via analysis of the Yb 4$f$ lineshape.

Fig. 2 presents the Fermi surface contours and reveals the dimensionality of the states at $\overline{\textmd{X}}$.
Panel 2a shows the near-E$_{\textmd{F}}$ energy dispersion acquired with $h\nu=70$ eV, an
excitation energy favorable for the states at $\overline{\textmd{X}}$ but not for those centered on $\overline{\Gamma}$.
The former yield elliptical Fermi contours (Fig. 2b), which decrease in size as the energy approaches the band bottom (Figs. 2b-2d).
Such elliptical contours around $\overline{\textmd{X}}$ are reminiscent of the elliptical contours observed in the band-bent version of other divalent hexaborides \cite{Denlinger2002,Denlinger2002_2} and have been attributed to the bulk conduction band.
Nevertheless, comparison of Figs. 2c and 2e shows that the states at $\overline{\textmd{X}}$ show no dispersion in the $k_\textmd{z}$ range covered, which suggests two-dimensional character: in the context of a cubic crystal such as YbB$_{6}$, this means surface-confined states.
The $k_\textmd{z}$ data presented in Fig. 2e were obtained by changing the photon energy, whereby the total $k_\textmd{z}$-range was kept modest so as to avoid  time-dependent changes of the electronic states which will be described in the context of Fig. 3.
In our experiments, we find the $\overline{\Gamma}$-pocket to only be clearly observable for a narrow range of photon energies between 30 and 35 eV. In this range also no $h\nu$-dependent changes of its dispersion were observed.

Thus, at this stage, the electron pockets at both $\overline{\Gamma}$ and $\overline{\textmd{X}}$ would appear to be confined in the near-surface region in agreement with Refs. \onlinecite{Feng2014, Ming2014, Hasan2014}. With respect to possible topological characteristics, the combined consideration of Figs. 1 and 2 leads to the conclusion that these states are electron-like, closed Fermi contours around both the $\overline{\Gamma}$ and $\overline{\textmd{X}}$ points, and that these come from two-dimensional states, but unlike Refs. \onlinecite{Feng2014, Ming2014}, our data lead us to conclude that these states do not possess the dispersion characteristics expected for edge states in a 3D topological insulator, correlated or otherwise.
Although we cannot exclude the possibility that these two-dimensional states correspond to true surface states of trivial character, the similarity of their ($k_{\textmd{x}}$,$k_{\textmd{y}}$) fingerprint with the contours of the bulk conduction band seen in other band-bent divalent hexaborides \cite{Denlinger2002,Denlinger2002_2} and the observation of two, concentric electron pockets around $\overline{\Gamma}$ (Fig. 1d) point towards a likely origin in quantum well states from confinement of the bulk conduction band.

In Fig. 3 we focus on a different characteristic which has not been reported in the recent ARPES studies of YbB$_6$ \cite{Feng2014, Ming2014, Hasan2014}, but has been reported previously for EuB$_6$ \cite{Denlinger2002,Denlinger2002_2}, namely that there is a pronounced time-dependence in the ARPES data, resulting in a gradual energy shift and modification of the surface-related features in YbB$_{6}$.

Fig. 3a presents a wide $k_{\|}$-range Fermi surface of YbB$_{6}$(001).
Red solid lines denote the borders of the 1$^{\textmd{st}}$ SBZ.
The $\overline{\Gamma}$ and $\overline{\textmd{X}}$ electron pockets form circular and elliptical contours respectively, and in these data there is no sign of the $2\times1$ backfolding reported in Ref. \onlinecite{Ming2014}.
The type of Fermi surface shown in Fig. 3a is typical for freshly cleaved YbB$_{6}$(001) samples. 
Closer examination of Fig. 3a reveals that the constant energy contours of all features decrease in size as one moves from the bottom part of the Fermi surface map to the top.
The acquisition of the map - as indicated in Fig. 3a - started 30 min after cleavage (bottom) and finished 210 min after cleavage (top).
The fact that there is a time-dependence in the data is also readily seen in Fig. 3b where the experimental band dispersion is presented along the $k$-space path highlighted by the green dashed line in Fig. 3a. It is clear that the electron pocket at $\overline{\textmd{X}}$ on the positive $k_{\|}$ side of Fig. 3b is shallower than at the symmetry-equivalent $\overline{\textmd{X}}$ point with negative $k_{\|}$ values: the band bottoms differing by nearly 50 meV.
A similar time-dependent energy shift to lower binding energies can be deduced for the  $\overline{\Gamma}$ pocket, if one compares the size of the two circular contours surrounding the labelled $\overline{\Gamma}$ points in Fig. 3a.

As a result of this time dependence, the provision of hard numbers for the energy position of band bottom and/or the $k_{\textmd{F}}$ values for either the $\overline{\textmd{X}}$ or the $\overline{\Gamma}$ electron pockets should involve the history of the cleave.
Fig. 3c illustrates this by following the time-dependent energy shift of the states at $\overline{\textmd{X}}$ systematically.
For freshly cleaved samples (20 min) the band minimum lies as deep as 200 meV below E$_{\textmd{F}}$, while it gradually shifts to a binding energy of 110 meV after 150 minutes.
We note that the I(E,$k$) images show negligible broadening as they shift to lower E$_{\textmd{B}}$. Thus the observed time-dependence of the electronic structure is not  an expression of a general degradation of the sample surface but rather indicates a rigid shift of well-defined bands. 
At longer timescales, the modification of the surface electronic structure continues and we found that after 4 hours the electron pocket at $\overline{\textmd{X}}$ is both shallow and ultimately very weak.
Beyond 5 hours after cleavage, neither the $\overline{\textmd{X}}$ nor the $\overline{\Gamma}$ electron pockets could be found on any part of the sample, indicating a global modification of the surface electronic structure which will be further discussed in the context of Fig. 4.
Such changes of the chemical potential may be attributed to a time-dependent modification of the initial surface band bending due to the adsorption of residual gas atoms and/or the effect of the photon beam. Similar changes of the initial surface band bending conditions have been observed on various compounds \cite{Meevasana2011,King2011} including divalent hexaborides \cite{Denlinger2002,Denlinger2002_2}. 

For completeness, we note here that the band bottom of the $\overline{\textmd{X}}$ electron pockets from fresh cleaves lies at slightly lower binding energy (0.19 eV) in these data than in those reported in other ARPES studies (0.25 - 0.3 eV) \cite{Feng2014, Ming2014, Hasan2014}.
Apart from the open issue of the history of the cleaves in each case, this difference could also be due to the higher temperature of our measurements (37K), as changing temperature has been shown to alter the energetics of the low-lying bands \cite{Hasan2014}.
\begin{figure}
  \centering
  \includegraphics[width = 8.7 cm]{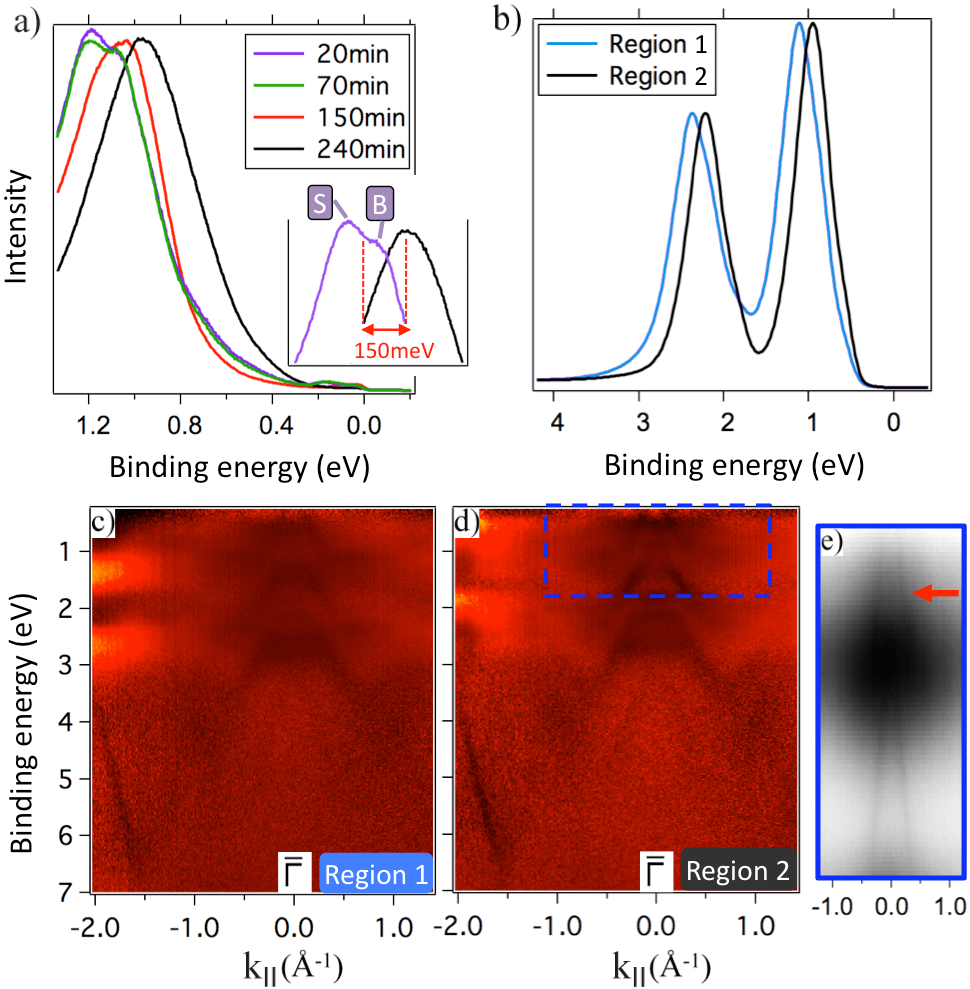}
  \caption{
(a) Angle-integrated ARPES intensity along $\overline{\textmd{X}\Gamma\textmd{X}}$, showing the time-evolution of the Yb 4$f$ states closest to E$_F$. The inset compares the double-peak lineshape of the Yb 4$f$ states measured on a fresh cleave (purple) with the lineshape measured after exposure to UHV (black). ``S'' and ``B'' denote the surface- and bulk-derived components of the higher binding energy feature, respectively. The time labels given refer to the time after cleavage.
(b) Spatial variations in the binding energy of the Yb 4$f$ states (data acquired 240 min after cleavage). Regions 1 and 2 are situated 100 $\mu$m away from each other.
(c),(d) Valence band data acquired from regions 1 and 2. 
In order to enhance the contribution from the dispersive features in panels (c) and (d), the data are normalized to have equal total intensity for each horizontal, momentum distribution slice.
(e) Raw data in the E($k$) region of highlighted in panel (d) by the dashed rectangle. The arrow points out the hole-like feature which
dominates the near-E$_{\textmd{F}}$ electronic structure of YbB$_{6}$(001) 240 min after cleavage.
The data of panel (a) were acquired with $h\nu=70$ eV, and (b)-(e) with $h\nu=150$ eV.
The sample temperature was 37K for all data shown.}
\label{Fig4}
\end{figure}

In Fig. 4 we report both spatial and temporal variations of the electronic structure of the YbB$_{6}$(001) surface.
Fig. 4a follows the time-dependent energy shift of the lowest lying Yb 4$f$ states.
The total shift of 150 meV between 20 min and 240 min after cleavage
compares well with the energy shift of the electron pocket seen at $\overline{\textmd{X}}$ over the same time-period (Fig. 3c), pointing again to a 
time-dependent rigid energy shift of the whole band structure at the surface.
Moreover, there is a clear double-peak structure for the Yb 4$f$ states on fresh surfaces which gradually loses definition to become a single, broad peak as the time after cleavage increases.
Following past work on the Yb 4$f$ line shape in YbB$_{6}$ \cite{Kakizaki1993} and other compounds \cite{Harada1999,Joyce1996}, we attribute the feature at
higher binding energy (labelled S in the inset to Fig. 4a) to divalent Yb atoms at the surface and the feature at lower binding energy (labelled B) to their bulk
counterparts.
The existence of Yb surface ions with a higher binding energy (feature S) means a more positive electrostatic potential at the surface than in the bulk. Such
a surface vs. bulk potential difference may give rise to the band bending situation described above in the context of Fig. 1, in which the formerly unoccupied conduction band shifts below the Fermi level.
A comparison of Figs. 4a and 3c reveals indeed that as the surface band bending seen in ARPES decreases, the intensity of the surface-related component in the Yb 4$f$ lineshape gradually decreases.   
This loss in intensity and ultimately definition of the surface-related 4$f$ feature as the time from cleavage progresses could result from the 
co-ordination of surface Yb atoms becoming more bulk-like due either to relaxation of the surface structure or due the effects of adsorbates on the surface. 

Data on the spatial homogeneity of the surface electronic structure is presented in Figs. 4b-d.
Fig. 4b presents an example of spatial variations on the electronic
structure of a sample which has been exposed to the UHV environment for 240 min.
The black curve (labelled region 2) shows the dominant energy position of the Yb 4$f$ spin-orbit doublet for this surface. 
However, on the same cleave, we could observe sample locations (labelled region 1) in which energy position of the 4$f$ states is similar to those of the bulk-like signal from a freshly cleaved sample.
The two regions probed in Fig. 4b were 100-200 $\mu$m away from each other, a distance which is comparable to the size of the synchrotron beam spot.
For the 240 minute time-point after cleavage relevant for Fig. 4b, at neither region 1 nor 2 could the high binding energy surface-derived component (S) be observed in the Yb 4$f$ lineshape, nor could the electron pockets at either $\overline{\textmd{X}}$ or $\overline{\Gamma}$ be seen in the ARPES data.

In Figs. 4c and 4d we plot the I(E,$k$) images for the valence bands of regions 1 and 2, whereby the dispersive features have been enhanced versus the flat 4$f$ states (see caption).
The dispersive bands mainly originate from B 2$p$ orbitals, as the boron framework defines the energy dispersion in metal hexaborides to a substantial
degree, with B 2$p$-related states dominating the electronic structure at binding energies higher than 1 eV \cite{Schmitt2001, Mackinnon2013, Alarco2014}. 
Noticeable differences between Figs. 4c and 4d are the rigid energy shift already shown in Fig. 4b and the fact that the B 2$p$ valence band states are more clearly resolved in panel (d) [region 2] than panel (c) [region 1].
This could mean that regions 1 and 2 present different terminations of the YbB$_{6}$ crystal.
Fig. 4e presents an expanded view of the raw valence band ARPES data for region 2 (see dashed rectangle in Fig. 4d).
Given the possibility of different surface terminations, it is conceivable that the hole-like feature highlighted with the red arrow in Fig. 4e could also be present on fresh cleaves for which the region 2 termination dominates.  
In Refs. \onlinecite{Feng2014} and \onlinecite{Ming2014}, $p$-type surfaces are mentioned, and are reported to possess hole-like features at E$_{\textmd{F}}$.
However, rather than being the bottom part of a Dirac cone dispersion, the data presented here could indicate that
these features from $p$-type surfaces may be the result of the (time-dependent) modification of the electronic structure. 
We note that the electron pockets at $\overline{\textmd{X}}$ / $\overline{\Gamma}$ and the hole-like features were never observed simultaneously
for the same YbB$_{6}$ surface region, neither in the data presented here nor in Refs. \onlinecite{Feng2014, Ming2014}. This is quite unlike
theory predictions for a Dirac cone in YbB$_{6}$ \cite{Dai2014} and unlike ARPES results from confirmed topological insulators (e.g. Ref. \onlinecite{Hasan2010} and references therein).\\ 

\section*{Conclusions}

We have identified two types of metallic states in the electronic band structure of UHV-cleaved YbB$_{6}$(001), which form closed Fermi contours centered around the $\overline{\textmd{X}}$ and $\overline{\Gamma}$-points of the surface Brillouin zone.
In agreement with previous studies, the photon energy dependence of these features shows them to be quasi-2D and thus confined in the near-surface region.
Unlike in previous studies, their ARPES signature is not masked either by other electronic states or matrix element effects, enabling the important observation that the dispersion relation for both states strongly resembles that of a parabola with a clearly identifiable band bottom. 
The absence of any ARPES feature in our data that could be argued to form the lower part of a Dirac cone points towards a topologically trivial character for these metallic states we observe at the (001) surface of cleaved YbB$_{6}$.
The simplest explanation is that this material is not a topological insulator, but it is conceivable that topological character may still be found for different - as yet unobserved- states or perhaps at different cleavage surfaces of YbB$_{6}$.

In addition, on exposure to the UHV environment, both sets of observed metallic states undergo a gradual energy shift towards lower binding
energies, and are no longer detectable some 4-5 hours after cleavage. By a combined consideration of time-dependent changes in the near-E$_{\textmd{F}}$ electronic structure and in the Yb 4$f$ lineshape, we propose a surface band bending scenario on the (001) cleavage surface of YbB$_{6}$ in line with past results on other divalent hexaborides.
The pronounced time dependence of the YbB$_{6}$(001) metallic states under UHV conditions is in contrast to the metallic states seen under identical experimental conditions in isostructural SmB$_{6}$(001), which are relatively insensitive to UHV aging.
Factoring in that the electronic structure of YbB$_{6}$ seen by ARPES also varies depending on the position chosen on the cleavage surface, leads to the overall conclusion that the electronic structure of the (001) cleavage surface of high quality single crystals differs significantly from recent theoretical expectations, and that the ARPES data presented here are unable to support the existence of topological surface states.\\      

\section*{Methods}

\subsection*{Sample growth and cleavage}YbB$_{6}$ single crystals were grown in an optical floating zone furnace
(Crystal Systems Inc. FZ-T-12000-S-BU-PC) under 5 MPa pressure of high purity argon gas \cite{Bao2013}. The growth rate was 20 mm per hour
with the feed and seed rods counter rotating at 30 rpm. Samples were cleaved at 37K at a pressure lower than 3.0$\times$10$^{-10}$ mbar.\\
\subsection*{Angle resolved photoelectron spectroscopy}
ARPES experiments were performed at the UE112-PGM-2a-1$^{2}$ beamline (BESSY II storage ring at the Helmholtz Zentrum Berlin) using a Scienta R8000 hemispherical electron analyzer and a six-axis manipulator. The pressure during measurements was 1.0$\times$10$^{-10}$ mbar, while the sample temperature was maintained at 37K. The energy position of the Fermi level was determined for every cleave by evaporation
of Au films onto the sample holder, in direct electrical contact with the crystal.
The polarization of the incoming photon beam was linear horizontal and the entrance slits of the hemispherical analyzer were vertical. The whole surface Brillouin zone was spanned by sequential rotation of the polar angle.\\

\section*{Acknowledgements}    
This work is part of the research programme of the Foundation for Fundamental Research on Matter (FOM), which is part of the Netherlands Organisation for Scientific Research (NWO). In addition, funding from the European Community's Seventh Framework Programme (FP7/2007-2013) under grant agreement 312284 is gratefully acknowledged, as are interesting discussions with Jonathan Denlinger, George Sawatzky and Ilya Elfimov.


\begin{thebibliography}{54}
\expandafter\ifx\csname natexlab\endcsname\relax\def\natexlab#1{#1}\fi
\expandafter\ifx\csname bibnamefont\endcsname\relax
  \def\bibnamefont#1{#1}\fi
\expandafter\ifx\csname bibfnamefont\endcsname\relax
  \def\bibfnamefont#1{#1}\fi
\expandafter\ifx\csname citenamefont\endcsname\relax
  \def\citenamefont#1{#1}\fi
\expandafter\ifx\csname url\endcsname\relax
  \def\url#1{\texttt{#1}}\fi
\expandafter\ifx\csname urlprefix\endcsname\relax\def\urlprefix{URL }\fi
\providecommand{\bibinfo}[2]{#2}
\providecommand{\eprint}[2][]{\url{#2}}

\bibitem[{\citenamefont{Hasan and Kane}(2010)}]{Hasan2010}
\bibinfo{author}{\bibfnamefont{M.~Z.} \bibnamefont{Hasan}} \bibnamefont{and}
  \bibinfo{author}{\bibfnamefont{C.~L.} \bibnamefont{Kane}},
  \bibinfo{journal}{Rev. Mod. Phys.} \textbf{\bibinfo{volume}{82}},
  \bibinfo{pages}{3045} (\bibinfo{year}{2010}).

\bibitem[{\citenamefont{Pesin and Balents}(2010)}]{Pesin2010}
\bibinfo{author}{\bibfnamefont{D.}~\bibnamefont{Pesin}} \bibnamefont{and}
  \bibinfo{author}{\bibfnamefont{L.}~\bibnamefont{Balents}},
  \bibinfo{journal}{Nat. Phys.} \textbf{\bibinfo{volume}{6}},
  \bibinfo{pages}{376} (\bibinfo{year}{2010}).

\bibitem[{\citenamefont{Sheng et~al.}(2014)\citenamefont{Sheng, Gu, Sun, and
  Sheng}}]{Sheng2011}
\bibinfo{author}{\bibfnamefont{D.~N.} \bibnamefont{Sheng}},
  \bibinfo{author}{\bibfnamefont{Z.-C.} \bibnamefont{Gu}},
  \bibinfo{author}{\bibfnamefont{K.}~\bibnamefont{Sun}}, \bibnamefont{and}
  \bibinfo{author}{\bibfnamefont{L.}~\bibnamefont{Sheng}},
  \bibinfo{journal}{Nat. Commun.} \textbf{\bibinfo{volume}{2}},
  \bibinfo{pages}{389} (\bibinfo{year}{2014}).

\bibitem[{\citenamefont{Regnault and Bernevig}(2011)}]{Regnault2011}
\bibinfo{author}{\bibfnamefont{N.}~\bibnamefont{Regnault}} \bibnamefont{and}
  \bibinfo{author}{\bibfnamefont{B.~A.} \bibnamefont{Bernevig}},
  \bibinfo{journal}{Phys. Rev. X} \textbf{\bibinfo{volume}{1}},
  \bibinfo{pages}{021014} (\bibinfo{year}{2011}).

\bibitem[{\citenamefont{Qia and Zhang}(2010)}]{Qi2011}
\bibinfo{author}{\bibfnamefont{X.-L.} \bibnamefont{Qia}} \bibnamefont{and}
  \bibinfo{author}{\bibfnamefont{S.-C.} \bibnamefont{Zhang}},
  \bibinfo{journal}{Rev. Mod. Phys.} \textbf{\bibinfo{volume}{83}},
  \bibinfo{pages}{1057} (\bibinfo{year}{2010}).

\bibitem[{\citenamefont{Dzero et~al.}(2010)\citenamefont{Dzero, Sun, Galitski,
  and Coleman}}]{Dzero2010}
\bibinfo{author}{\bibfnamefont{M.}~\bibnamefont{Dzero}},
  \bibinfo{author}{\bibfnamefont{K.}~\bibnamefont{Sun}},
  \bibinfo{author}{\bibfnamefont{V.}~\bibnamefont{Galitski}}, \bibnamefont{and}
  \bibinfo{author}{\bibfnamefont{P.}~\bibnamefont{Coleman}},
  \bibinfo{journal}{Phys. Rev. Lett.} \textbf{\bibinfo{volume}{104}},
  \bibinfo{pages}{106408} (\bibinfo{year}{2010}).

\bibitem[{\citenamefont{Alexandrov et~al.}(2013)\citenamefont{Alexandrov,
  Dzero, and Coleman}}]{Alexandrov2013}
\bibinfo{author}{\bibfnamefont{V.}~\bibnamefont{Alexandrov}},
  \bibinfo{author}{\bibfnamefont{M.}~\bibnamefont{Dzero}}, \bibnamefont{and}
  \bibinfo{author}{\bibfnamefont{P.}~\bibnamefont{Coleman}},
  \bibinfo{journal}{Phys. Rev. Lett.} \textbf{\bibinfo{volume}{111}},
  \bibinfo{pages}{226403} (\bibinfo{year}{2013}).

\bibitem[{\citenamefont{Lu et~al.}(2013)\citenamefont{Lu, Zhao, Weng, Fang, and
  Dai}}]{Lu2013}
\bibinfo{author}{\bibfnamefont{F.}~\bibnamefont{Lu}},
  \bibinfo{author}{\bibfnamefont{J.}~\bibnamefont{Zhao}},
  \bibinfo{author}{\bibfnamefont{H.}~\bibnamefont{Weng}},
  \bibinfo{author}{\bibfnamefont{Z.}~\bibnamefont{Fang}}, \bibnamefont{and}
  \bibinfo{author}{\bibfnamefont{X.}~\bibnamefont{Dai}},
  \bibinfo{journal}{Phys. Rev. Lett.} \textbf{\bibinfo{volume}{110}},
  \bibinfo{pages}{096401} (\bibinfo{year}{2013}).

\bibitem[{\citenamefont{Wolgast et~al.}(2013)\citenamefont{Wolgast, Kurdak,
  Sun, Allen, Kim, and Fisk}}]{Wolgast2013}
\bibinfo{author}{\bibfnamefont{S.}~\bibnamefont{Wolgast}},
  \bibinfo{author}{\bibfnamefont{C.}~\bibnamefont{Kurdak}},
  \bibinfo{author}{\bibfnamefont{K.}~\bibnamefont{Sun}},
  \bibinfo{author}{\bibfnamefont{J.~W.} \bibnamefont{Allen}},
  \bibinfo{author}{\bibfnamefont{D.-J.} \bibnamefont{Kim}}, \bibnamefont{and}
  \bibinfo{author}{\bibfnamefont{Z.}~\bibnamefont{Fisk}},
  \bibinfo{journal}{Phys. Rev. B} \textbf{\bibinfo{volume}{88}},
  \bibinfo{pages}{180405} (\bibinfo{year}{2013}).

\bibitem[{\citenamefont{Li et~al.}()\citenamefont{Li, Xiang, Yu, Asab, Lawson,
  Cai, Tinsman, Berkley, Wolgast, Eo et~al.}}]{Li2013}
\bibinfo{author}{\bibfnamefont{G.}~\bibnamefont{Li}},
  \bibinfo{author}{\bibfnamefont{Z.}~\bibnamefont{Xiang}},
  \bibinfo{author}{\bibfnamefont{F.}~\bibnamefont{Yu}},
  \bibinfo{author}{\bibfnamefont{T.}~\bibnamefont{Asab}},
  \bibinfo{author}{\bibfnamefont{B.}~\bibnamefont{Lawson}},
  \bibinfo{author}{\bibfnamefont{P.}~\bibnamefont{Cai}},
  \bibinfo{author}{\bibfnamefont{C.}~\bibnamefont{Tinsman}},
  \bibinfo{author}{\bibfnamefont{A.}~\bibnamefont{Berkley}},
  \bibinfo{author}{\bibfnamefont{S.}~\bibnamefont{Wolgast}},
  \bibinfo{author}{\bibfnamefont{Y.~S.} \bibnamefont{Eo}},
  \bibnamefont{et~al.}, \bibinfo{note}{arXiv:1306.5221 (2013)}.

\bibitem[{\citenamefont{Kim et~al.}(2013)\citenamefont{Kim, Thomas, Grant,
  Botimer, Fisk, and Xia}}]{Fisk2013}
\bibinfo{author}{\bibfnamefont{D.~J.} \bibnamefont{Kim}},
  \bibinfo{author}{\bibfnamefont{S.}~\bibnamefont{Thomas}},
  \bibinfo{author}{\bibfnamefont{T.}~\bibnamefont{Grant}},
  \bibinfo{author}{\bibfnamefont{J.}~\bibnamefont{Botimer}},
  \bibinfo{author}{\bibfnamefont{Z.}~\bibnamefont{Fisk}}, \bibnamefont{and}
  \bibinfo{author}{\bibfnamefont{J.}~\bibnamefont{Xia}}, \bibinfo{journal}{Sci.
  Rep.} \textbf{\bibinfo{volume}{3}}, \bibinfo{pages}{3150}
  (\bibinfo{year}{2013}).

\bibitem[{\citenamefont{Zhang et~al.}(2013)\citenamefont{Zhang, Butch, Syers,
  Ziemak, Greene, and Paglione}}]{Zhang2013}
\bibinfo{author}{\bibfnamefont{X.}~\bibnamefont{Zhang}},
  \bibinfo{author}{\bibfnamefont{N.~P.} \bibnamefont{Butch}},
  \bibinfo{author}{\bibfnamefont{P.}~\bibnamefont{Syers}},
  \bibinfo{author}{\bibfnamefont{S.}~\bibnamefont{Ziemak}},
  \bibinfo{author}{\bibfnamefont{R.~L.} \bibnamefont{Greene}},
  \bibnamefont{and} \bibinfo{author}{\bibfnamefont{J.}~\bibnamefont{Paglione}},
  \bibinfo{journal}{Phys. Rev. X} \textbf{\bibinfo{volume}{3}},
  \bibinfo{pages}{011011} (\bibinfo{year}{2013}).

\bibitem[{\citenamefont{Kim et~al.}(2014)\citenamefont{Kim, Xia, and
  Fisk}}]{Fisk2014}
\bibinfo{author}{\bibfnamefont{D.~J.} \bibnamefont{Kim}},
  \bibinfo{author}{\bibfnamefont{J.}~\bibnamefont{Xia}}, \bibnamefont{and}
  \bibinfo{author}{\bibfnamefont{Z.}~\bibnamefont{Fisk}},
  \bibinfo{journal}{Nat. Mater.} \textbf{\bibinfo{volume}{13}},
  \bibinfo{pages}{466} (\bibinfo{year}{2014}).

\bibitem[{\citenamefont{Neupane et~al.}(2013)\citenamefont{Neupane, Alidoust,
  Xu, Kondo, Ishida, Kim, Liu, Belopolski, Jo, Chang et~al.}}]{Hasan2013}
\bibinfo{author}{\bibfnamefont{M.}~\bibnamefont{Neupane}},
  \bibinfo{author}{\bibfnamefont{A.}~\bibnamefont{Alidoust}},
  \bibinfo{author}{\bibfnamefont{S.-Y.} \bibnamefont{Xu}},
  \bibinfo{author}{\bibfnamefont{T.}~\bibnamefont{Kondo}},
  \bibinfo{author}{\bibfnamefont{Y.}~\bibnamefont{Ishida}},
  \bibinfo{author}{\bibfnamefont{D.~J.} \bibnamefont{Kim}},
  \bibinfo{author}{\bibfnamefont{C.}~\bibnamefont{Liu}},
  \bibinfo{author}{\bibfnamefont{I.}~\bibnamefont{Belopolski}},
  \bibinfo{author}{\bibfnamefont{Y.~J.} \bibnamefont{Jo}},
  \bibinfo{author}{\bibfnamefont{T.-R.} \bibnamefont{Chang}},
  \bibnamefont{et~al.}, \bibinfo{journal}{Nat. Commun.}
  \textbf{\bibinfo{volume}{4}}, \bibinfo{pages}{2991} (\bibinfo{year}{2013}).

\bibitem[{\citenamefont{Xu et~al.}(2013)\citenamefont{Xu, Shi, Biswas, Matt,
  Dhaka, Huang, Plumb, Radovic, Dil, Pomjakushina et~al.}}]{Ming2013}
\bibinfo{author}{\bibfnamefont{N.}~\bibnamefont{Xu}},
  \bibinfo{author}{\bibfnamefont{X.}~\bibnamefont{Shi}},
  \bibinfo{author}{\bibfnamefont{P.~K.} \bibnamefont{Biswas}},
  \bibinfo{author}{\bibfnamefont{C.~E.} \bibnamefont{Matt}},
  \bibinfo{author}{\bibfnamefont{R.~S.} \bibnamefont{Dhaka}},
  \bibinfo{author}{\bibfnamefont{Y.}~\bibnamefont{Huang}},
  \bibinfo{author}{\bibfnamefont{N.~C.} \bibnamefont{Plumb}},
  \bibinfo{author}{\bibfnamefont{M.}~\bibnamefont{Radovic}},
  \bibinfo{author}{\bibfnamefont{J.~H.} \bibnamefont{Dil}},
  \bibinfo{author}{\bibfnamefont{E.}~\bibnamefont{Pomjakushina}},
  \bibnamefont{et~al.}, \bibinfo{journal}{Phys. Rev. B}
  \textbf{\bibinfo{volume}{88}}, \bibinfo{pages}{121102}
  (\bibinfo{year}{2013}).

\bibitem[{\citenamefont{Jiang et~al.}(2013)\citenamefont{Jiang, Li, Zhang, Sun,
  Chen, Ye, Xu, Ge, Tan, Niu et~al.}}]{Feng2013}
\bibinfo{author}{\bibfnamefont{J.}~\bibnamefont{Jiang}},
  \bibinfo{author}{\bibfnamefont{S.}~\bibnamefont{Li}},
  \bibinfo{author}{\bibfnamefont{T.}~\bibnamefont{Zhang}},
  \bibinfo{author}{\bibfnamefont{Z.}~\bibnamefont{Sun}},
  \bibinfo{author}{\bibfnamefont{F.}~\bibnamefont{Chen}},
  \bibinfo{author}{\bibfnamefont{Z.~R.} \bibnamefont{Ye}},
  \bibinfo{author}{\bibfnamefont{M.}~\bibnamefont{Xu}},
  \bibinfo{author}{\bibfnamefont{Q.~Q.} \bibnamefont{Ge}},
  \bibinfo{author}{\bibfnamefont{S.~Y.} \bibnamefont{Tan}},
  \bibinfo{author}{\bibfnamefont{X.~H.} \bibnamefont{Niu}},
  \bibnamefont{et~al.}, \bibinfo{journal}{Nat. Commun.}
  \textbf{\bibinfo{volume}{4}}, \bibinfo{pages}{3010} (\bibinfo{year}{2013}).

\bibitem[{\citenamefont{Frantzeskakis et~al.}(2013)\citenamefont{Frantzeskakis,
  de~Jong, Zwartsenberg, Huang, Pan, Zhang, Zhang, Zhang, Bao, Tegus
  et~al.}}]{Frantzeskakis2013}
\bibinfo{author}{\bibfnamefont{E.}~\bibnamefont{Frantzeskakis}},
  \bibinfo{author}{\bibfnamefont{N.}~\bibnamefont{de~Jong}},
  \bibinfo{author}{\bibfnamefont{B.}~\bibnamefont{Zwartsenberg}},
  \bibinfo{author}{\bibfnamefont{Y.~K.} \bibnamefont{Huang}},
  \bibinfo{author}{\bibfnamefont{Y.}~\bibnamefont{Pan}},
  \bibinfo{author}{\bibfnamefont{X.}~\bibnamefont{Zhang}},
  \bibinfo{author}{\bibfnamefont{J.~X.} \bibnamefont{Zhang}},
  \bibinfo{author}{\bibfnamefont{F.~X.} \bibnamefont{Zhang}},
  \bibinfo{author}{\bibfnamefont{L.~H.} \bibnamefont{Bao}},
  \bibinfo{author}{\bibfnamefont{O.}~\bibnamefont{Tegus}},
  \bibnamefont{et~al.}, \bibinfo{journal}{Phys. Rev. X}
  \textbf{\bibinfo{volume}{3}}, \bibinfo{pages}{041024} (\bibinfo{year}{2013}).

\bibitem[{\citenamefont{Zhu et~al.}(2013)\citenamefont{Zhu, Nicolaou, Levy,
  Butch, Syers, Wang, Paglione, Sawatzky, Elfimov, and
  Damascelli}}]{Damascelli2013}
\bibinfo{author}{\bibfnamefont{Z.-H.} \bibnamefont{Zhu}},
  \bibinfo{author}{\bibfnamefont{A.}~\bibnamefont{Nicolaou}},
  \bibinfo{author}{\bibfnamefont{G.}~\bibnamefont{Levy}},
  \bibinfo{author}{\bibfnamefont{N.~P.} \bibnamefont{Butch}},
  \bibinfo{author}{\bibfnamefont{P.}~\bibnamefont{Syers}},
  \bibinfo{author}{\bibfnamefont{X.~F.} \bibnamefont{Wang}},
  \bibinfo{author}{\bibfnamefont{J.}~\bibnamefont{Paglione}},
  \bibinfo{author}{\bibfnamefont{G.~A.} \bibnamefont{Sawatzky}},
  \bibinfo{author}{\bibfnamefont{I.~S.} \bibnamefont{Elfimov}},
  \bibnamefont{and}
  \bibinfo{author}{\bibfnamefont{A.}~\bibnamefont{Damascelli}},
  \bibinfo{journal}{Phys. Rev. Lett.} \textbf{\bibinfo{volume}{111}},
  \bibinfo{pages}{216402} (\bibinfo{year}{2013}).

\bibitem[{\citenamefont{Denlinger et~al.}()\citenamefont{Denlinger, Allen,
  Kang, Sun, Kim, Shim, Min, Kim, and Fisk}}]{Denlinger2013}
\bibinfo{author}{\bibfnamefont{J.~D.} \bibnamefont{Denlinger}},
  \bibinfo{author}{\bibfnamefont{J.~W.} \bibnamefont{Allen}},
  \bibinfo{author}{\bibfnamefont{J.-S.} \bibnamefont{Kang}},
  \bibinfo{author}{\bibfnamefont{K.}~\bibnamefont{Sun}},
  \bibinfo{author}{\bibfnamefont{J.-W.} \bibnamefont{Kim}},
  \bibinfo{author}{\bibfnamefont{J.~H.} \bibnamefont{Shim}},
  \bibinfo{author}{\bibfnamefont{B.~I.} \bibnamefont{Min}},
  \bibinfo{author}{\bibfnamefont{D.-J.} \bibnamefont{Kim}}, \bibnamefont{and}
  \bibinfo{author}{\bibfnamefont{Z.}~\bibnamefont{Fisk}},
  \bibinfo{note}{arXiv:1312.6637 (2013)}.

\bibitem[{\citenamefont{Min et~al.}(2014)\citenamefont{Min, Lutz, Fiedler,
  Kang, Cho, Kim, Bentmann, and Reinert}}]{Reinert2014}
\bibinfo{author}{\bibfnamefont{C.-H.} \bibnamefont{Min}},
  \bibinfo{author}{\bibfnamefont{P.}~\bibnamefont{Lutz}},
  \bibinfo{author}{\bibfnamefont{S.}~\bibnamefont{Fiedler}},
  \bibinfo{author}{\bibfnamefont{B.~Y.} \bibnamefont{Kang}},
  \bibinfo{author}{\bibfnamefont{B.~K.} \bibnamefont{Cho}},
  \bibinfo{author}{\bibfnamefont{H.-D.} \bibnamefont{Kim}},
  \bibinfo{author}{\bibfnamefont{H.}~\bibnamefont{Bentmann}}, \bibnamefont{and}
  \bibinfo{author}{\bibfnamefont{F.}~\bibnamefont{Reinert}},
  \bibinfo{journal}{Phys. Rev. Lett.} \textbf{\bibinfo{volume}{112}},
  \bibinfo{pages}{226402} (\bibinfo{year}{2014}).

\bibitem[{\citenamefont{Yee et~al.}()\citenamefont{Yee, He, and. D.-J.~Kim,
  Fisk, and Hoffman}}]{Hoffman2013}
\bibinfo{author}{\bibfnamefont{M.~M.} \bibnamefont{Yee}},
  \bibinfo{author}{\bibfnamefont{Y.}~\bibnamefont{He}},
  \bibinfo{author}{\bibfnamefont{A.~S.} \bibnamefont{and. D.-J.~Kim}},
  \bibinfo{author}{\bibfnamefont{Z.}~\bibnamefont{Fisk}}, \bibnamefont{and}
  \bibinfo{author}{\bibfnamefont{J.~E.} \bibnamefont{Hoffman}},
  \bibinfo{note}{arXiv:1308.1058 (2013)}.

\bibitem[{\citenamefont{R\"{o}ssler et~al.}(2014)\citenamefont{R\"{o}ssler,
  Jang, Kim, Tjeng, Fisk, Steglich, and Wirth}}]{Rossler2014}
\bibinfo{author}{\bibfnamefont{S.}~\bibnamefont{R\"{o}ssler}},
  \bibinfo{author}{\bibfnamefont{T.-H.} \bibnamefont{Jang}},
  \bibinfo{author}{\bibfnamefont{D.~J.} \bibnamefont{Kim}},
  \bibinfo{author}{\bibfnamefont{L.~H.} \bibnamefont{Tjeng}},
  \bibinfo{author}{\bibfnamefont{Z.}~\bibnamefont{Fisk}},
  \bibinfo{author}{\bibfnamefont{F.}~\bibnamefont{Steglich}}, \bibnamefont{and}
  \bibinfo{author}{\bibfnamefont{S.}~\bibnamefont{Wirth}},
  \bibinfo{journal}{Proc. Nat. Acad. Sci.} \textbf{\bibinfo{volume}{111}},
  \bibinfo{pages}{4798} (\bibinfo{year}{2014}).

\bibitem[{\citenamefont{Ruan et~al.}(2014)\citenamefont{Ruan, Ye, Guo, Chen,
  Chen, Zhang, and Wang}}]{Ruan2014}
\bibinfo{author}{\bibfnamefont{W.}~\bibnamefont{Ruan}},
  \bibinfo{author}{\bibfnamefont{C.}~\bibnamefont{Ye}},
  \bibinfo{author}{\bibfnamefont{M.}~\bibnamefont{Guo}},
  \bibinfo{author}{\bibfnamefont{F.}~\bibnamefont{Chen}},
  \bibinfo{author}{\bibfnamefont{X.}~\bibnamefont{Chen}},
  \bibinfo{author}{\bibfnamefont{G.-M.} \bibnamefont{Zhang}}, \bibnamefont{and}
  \bibinfo{author}{\bibfnamefont{Y.}~\bibnamefont{Wang}},
  \bibinfo{journal}{Phys. Rev. Lett.} \textbf{\bibinfo{volume}{112}},
  \bibinfo{pages}{136401} (\bibinfo{year}{2014}).

\bibitem[{\citenamefont{Roushan et~al.}(2009)\citenamefont{Roushan, Seo,
  Parker, Hor, Hsieh, Qian, Richardella, Hasan, Cava, and
  Yazdani}}]{Roushan2009}
\bibinfo{author}{\bibfnamefont{P.}~\bibnamefont{Roushan}},
  \bibinfo{author}{\bibfnamefont{J.}~\bibnamefont{Seo}},
  \bibinfo{author}{\bibfnamefont{C.~V.} \bibnamefont{Parker}},
  \bibinfo{author}{\bibfnamefont{Y.~S.} \bibnamefont{Hor}},
  \bibinfo{author}{\bibfnamefont{D.}~\bibnamefont{Hsieh}},
  \bibinfo{author}{\bibfnamefont{D.}~\bibnamefont{Qian}},
  \bibinfo{author}{\bibfnamefont{A.}~\bibnamefont{Richardella}},
  \bibinfo{author}{\bibfnamefont{M.~Z.} \bibnamefont{Hasan}},
  \bibinfo{author}{\bibfnamefont{R.~J.} \bibnamefont{Cava}}, \bibnamefont{and}
  \bibinfo{author}{\bibfnamefont{A.}~\bibnamefont{Yazdani}},
  \bibinfo{journal}{Nature (London)} \textbf{\bibinfo{volume}{460}},
  \bibinfo{pages}{1106} (\bibinfo{year}{2009}).

\bibitem[{\citenamefont{Zhang et~al.}(2009)\citenamefont{Zhang, Cheng, Chen,
  Jian, Ma, He, Wang, Zhang, Dai, Fang et~al.}}]{Zhang2009}
\bibinfo{author}{\bibfnamefont{T.}~\bibnamefont{Zhang}},
  \bibinfo{author}{\bibfnamefont{P.}~\bibnamefont{Cheng}},
  \bibinfo{author}{\bibfnamefont{X.}~\bibnamefont{Chen}},
  \bibinfo{author}{\bibfnamefont{J.-F.} \bibnamefont{Jian}},
  \bibinfo{author}{\bibfnamefont{X.}~\bibnamefont{Ma}},
  \bibinfo{author}{\bibfnamefont{K.}~\bibnamefont{He}},
  \bibinfo{author}{\bibfnamefont{L.}~\bibnamefont{Wang}},
  \bibinfo{author}{\bibfnamefont{H.}~\bibnamefont{Zhang}},
  \bibinfo{author}{\bibfnamefont{X.}~\bibnamefont{Dai}},
  \bibinfo{author}{\bibfnamefont{Z.}~\bibnamefont{Fang}}, \bibnamefont{et~al.},
  \bibinfo{journal}{Phys. Rev. Lett.} \textbf{\bibinfo{volume}{103}},
  \bibinfo{pages}{266803} (\bibinfo{year}{2009}).

\bibitem[{\citenamefont{Alpichshev et~al.}(2010)\citenamefont{Alpichshev,
  Analytis, Chu, Fisher, Chen, Shen, Fang, and Kapitulnik}}]{Alpichshev2010}
\bibinfo{author}{\bibfnamefont{Z.}~\bibnamefont{Alpichshev}},
  \bibinfo{author}{\bibfnamefont{J.~G.} \bibnamefont{Analytis}},
  \bibinfo{author}{\bibfnamefont{J.-H.} \bibnamefont{Chu}},
  \bibinfo{author}{\bibfnamefont{I.~R.} \bibnamefont{Fisher}},
  \bibinfo{author}{\bibfnamefont{Y.~L.} \bibnamefont{Chen}},
  \bibinfo{author}{\bibfnamefont{Z.~X.} \bibnamefont{Shen}},
  \bibinfo{author}{\bibfnamefont{A.}~\bibnamefont{Fang}}, \bibnamefont{and}
  \bibinfo{author}{\bibfnamefont{A.}~\bibnamefont{Kapitulnik}},
  \bibinfo{journal}{Phys. Rev. Lett.} \textbf{\bibinfo{volume}{104}},
  \bibinfo{pages}{016401} (\bibinfo{year}{2010}).

\bibitem[{\citenamefont{Beidenkopf et~al.}(2011)\citenamefont{Beidenkopf,
  Roushan, Seo, Gorman, Drozdov, Hor, Cava, and Yazdani}}]{Beidenkopf2011}
\bibinfo{author}{\bibfnamefont{H.}~\bibnamefont{Beidenkopf}},
  \bibinfo{author}{\bibfnamefont{P.}~\bibnamefont{Roushan}},
  \bibinfo{author}{\bibfnamefont{J.}~\bibnamefont{Seo}},
  \bibinfo{author}{\bibfnamefont{L.}~\bibnamefont{Gorman}},
  \bibinfo{author}{\bibfnamefont{I.}~\bibnamefont{Drozdov}},
  \bibinfo{author}{\bibfnamefont{Y.~S.} \bibnamefont{Hor}},
  \bibinfo{author}{\bibfnamefont{R.~J.} \bibnamefont{Cava}}, \bibnamefont{and}
  \bibinfo{author}{\bibfnamefont{A.}~\bibnamefont{Yazdani}},
  \bibinfo{journal}{Nature Phys.} \textbf{\bibinfo{volume}{7}},
  \bibinfo{pages}{939} (\bibinfo{year}{2011}).

\bibitem[{\citenamefont{Weng et~al.}(2014)\citenamefont{Weng, Zhao, Wang, Fang,
  and Dai}}]{Dai2014}
\bibinfo{author}{\bibfnamefont{H.}~\bibnamefont{Weng}},
  \bibinfo{author}{\bibfnamefont{J.}~\bibnamefont{Zhao}},
  \bibinfo{author}{\bibfnamefont{Z.}~\bibnamefont{Wang}},
  \bibinfo{author}{\bibfnamefont{Z.}~\bibnamefont{Fang}}, \bibnamefont{and}
  \bibinfo{author}{\bibfnamefont{X.}~\bibnamefont{Dai}},
  \bibinfo{journal}{Phys. Rev. Lett.} \textbf{\bibinfo{volume}{112}},
  \bibinfo{pages}{016403} (\bibinfo{year}{2014}).

\bibitem[{\citenamefont{Aprea et~al.}(2010)\citenamefont{Aprea, Maspero,
  Masciocchi, Guagliardi, Albisetti, and Giunchi}}]{Aprea2013}
\bibinfo{author}{\bibfnamefont{A.}~\bibnamefont{Aprea}},
  \bibinfo{author}{\bibfnamefont{A.}~\bibnamefont{Maspero}},
  \bibinfo{author}{\bibfnamefont{N.}~\bibnamefont{Masciocchi}},
  \bibinfo{author}{\bibfnamefont{A.}~\bibnamefont{Guagliardi}},
  \bibinfo{author}{\bibfnamefont{A.~F.} \bibnamefont{Albisetti}},
  \bibnamefont{and} \bibinfo{author}{\bibfnamefont{G.}~\bibnamefont{Giunchi}},
  \bibinfo{journal}{Solid State Sci.} \textbf{\bibinfo{volume}{21}},
  \bibinfo{pages}{32} (\bibinfo{year}{2010}).

\bibitem[{\citenamefont{Tarascon et~al.}(1980)\citenamefont{Tarascon,
  Etourneau, Dordor, Hagenmuller, Kasaya, and Coey}}]{Tarascon1980}
\bibinfo{author}{\bibfnamefont{J.~M.} \bibnamefont{Tarascon}},
  \bibinfo{author}{\bibfnamefont{J.}~\bibnamefont{Etourneau}},
  \bibinfo{author}{\bibfnamefont{P.}~\bibnamefont{Dordor}},
  \bibinfo{author}{\bibfnamefont{P.}~\bibnamefont{Hagenmuller}},
  \bibinfo{author}{\bibfnamefont{M.}~\bibnamefont{Kasaya}}, \bibnamefont{and}
  \bibinfo{author}{\bibfnamefont{J.~M.~D.} \bibnamefont{Coey}},
  \bibinfo{journal}{J. Appl. Phys.} \textbf{\bibinfo{volume}{51}},
  \bibinfo{pages}{574} (\bibinfo{year}{1980}).

\bibitem[{\citenamefont{Nanba et~al.}(1993)\citenamefont{Nanba, Tomikawa, Mori,
  Shino, Imada, Suga, Kimura, and Kunii}}]{Nanba1993}
\bibinfo{author}{\bibfnamefont{T.}~\bibnamefont{Nanba}},
  \bibinfo{author}{\bibfnamefont{M.}~\bibnamefont{Tomikawa}},
  \bibinfo{author}{\bibfnamefont{Y.}~\bibnamefont{Mori}},
  \bibinfo{author}{\bibfnamefont{N.}~\bibnamefont{Shino}},
  \bibinfo{author}{\bibfnamefont{S.}~\bibnamefont{Imada}},
  \bibinfo{author}{\bibfnamefont{S.}~\bibnamefont{Suga}},
  \bibinfo{author}{\bibfnamefont{S.}~\bibnamefont{Kimura}}, \bibnamefont{and}
  \bibinfo{author}{\bibfnamefont{S.}~\bibnamefont{Kunii}},
  \bibinfo{journal}{Physica B} \textbf{\bibinfo{volume}{186-188}},
  \bibinfo{pages}{557} (\bibinfo{year}{1993}).

\bibitem[{\citenamefont{Young et~al.}(1999)\citenamefont{Young, Hall, Torelli,
  Fisk, Sarrao, Thompson, Ott, Oseroff, Goodrich, and Zysler}}]{Young1999}
\bibinfo{author}{\bibfnamefont{D.~P.} \bibnamefont{Young}},
  \bibinfo{author}{\bibfnamefont{D.}~\bibnamefont{Hall}},
  \bibinfo{author}{\bibfnamefont{M.~E.} \bibnamefont{Torelli}},
  \bibinfo{author}{\bibfnamefont{Z.}~\bibnamefont{Fisk}},
  \bibinfo{author}{\bibfnamefont{J.~L.} \bibnamefont{Sarrao}},
  \bibinfo{author}{\bibfnamefont{J.~D.} \bibnamefont{Thompson}},
  \bibinfo{author}{\bibfnamefont{H.-R.} \bibnamefont{Ott}},
  \bibinfo{author}{\bibfnamefont{S.~B.} \bibnamefont{Oseroff}},
  \bibinfo{author}{\bibfnamefont{R.~G.} \bibnamefont{Goodrich}},
  \bibnamefont{and} \bibinfo{author}{\bibfnamefont{R.}~\bibnamefont{Zysler}},
  \bibinfo{journal}{Nature (London)} \textbf{\bibinfo{volume}{397}},
  \bibinfo{pages}{412} (\bibinfo{year}{1999}).

\bibitem[{\citenamefont{Zhitomirsky et~al.}(1999)\citenamefont{Zhitomirsky,
  Rice, and Anisimov}}]{Zhitomirsky1999}
\bibinfo{author}{\bibfnamefont{M.~E.} \bibnamefont{Zhitomirsky}},
  \bibinfo{author}{\bibfnamefont{T.}~\bibnamefont{Rice}}, \bibnamefont{and}
  \bibinfo{author}{\bibfnamefont{V.~I.} \bibnamefont{Anisimov}},
  \bibinfo{journal}{Nature (London)} \textbf{\bibinfo{volume}{402}},
  \bibinfo{pages}{251} (\bibinfo{year}{1999}).

\bibitem[{\citenamefont{Balents and Varma}(2000)}]{Balents2000}
\bibinfo{author}{\bibfnamefont{L.}~\bibnamefont{Balents}} \bibnamefont{and}
  \bibinfo{author}{\bibfnamefont{C.~M.} \bibnamefont{Varma}},
  \bibinfo{journal}{Phys. Rev. Lett.} \textbf{\bibinfo{volume}{84}},
  \bibinfo{pages}{1264} (\bibinfo{year}{2000}).

\bibitem[{\citenamefont{Barzykin and Gor'kov}(2000)}]{Barzykin2000}
\bibinfo{author}{\bibfnamefont{V.}~\bibnamefont{Barzykin}} \bibnamefont{and}
  \bibinfo{author}{\bibfnamefont{L.~P.} \bibnamefont{Gor'kov}},
  \bibinfo{journal}{Phys. Rev. Lett.} \textbf{\bibinfo{volume}{84}},
  \bibinfo{pages}{2207} (\bibinfo{year}{2000}).

\bibitem[{\citenamefont{Murakami et~al.}(2002)\citenamefont{Murakami, Shindou,
  Nagaosa, and Mishchenko}}]{Murakami2002}
\bibinfo{author}{\bibfnamefont{S.}~\bibnamefont{Murakami}},
  \bibinfo{author}{\bibfnamefont{R.}~\bibnamefont{Shindou}},
  \bibinfo{author}{\bibfnamefont{N.}~\bibnamefont{Nagaosa}}, \bibnamefont{and}
  \bibinfo{author}{\bibfnamefont{A.~S.} \bibnamefont{Mishchenko}},
  \bibinfo{journal}{Phys. Rev. B} \textbf{\bibinfo{volume}{66}},
  \bibinfo{pages}{184405} (\bibinfo{year}{2002}).

\bibitem[{\citenamefont{Denlinger
  et~al.}(2002{\natexlab{a}})\citenamefont{Denlinger, Gweon, Mo, Allen, Sarrao,
  Bianchi, and Fisk}}]{Denlinger2002_2}
\bibinfo{author}{\bibfnamefont{J.~D.} \bibnamefont{Denlinger}},
  \bibinfo{author}{\bibfnamefont{G.-H.} \bibnamefont{Gweon}},
  \bibinfo{author}{\bibfnamefont{S.-K.} \bibnamefont{Mo}},
  \bibinfo{author}{\bibfnamefont{J.~W.} \bibnamefont{Allen}},
  \bibinfo{author}{\bibfnamefont{J.~L.} \bibnamefont{Sarrao}},
  \bibinfo{author}{\bibfnamefont{A.~D.} \bibnamefont{Bianchi}},
  \bibnamefont{and} \bibinfo{author}{\bibfnamefont{Z.}~\bibnamefont{Fisk}},
  \bibinfo{journal}{J. Phys. Soc. Jpn.} \textbf{\bibinfo{volume}{71}},
  \bibinfo{pages}{1} (\bibinfo{year}{2002}{\natexlab{a}}).

\bibitem[{\citenamefont{Kakizaki et~al.}(1993)\citenamefont{Kakizaki, Harasawa,
  Kinoshita, Ishii, Nanba, and Kunii}}]{Kakizaki1993}
\bibinfo{author}{\bibfnamefont{A.}~\bibnamefont{Kakizaki}},
  \bibinfo{author}{\bibfnamefont{A.}~\bibnamefont{Harasawa}},
  \bibinfo{author}{\bibfnamefont{T.}~\bibnamefont{Kinoshita}},
  \bibinfo{author}{\bibfnamefont{T.}~\bibnamefont{Ishii}},
  \bibinfo{author}{\bibfnamefont{T.}~\bibnamefont{Nanba}}, \bibnamefont{and}
  \bibinfo{author}{\bibfnamefont{S.}~\bibnamefont{Kunii}},
  \bibinfo{journal}{Physica B} \textbf{\bibinfo{volume}{186}},
  \bibinfo{pages}{80} (\bibinfo{year}{1993}).

\bibitem[{\citenamefont{Curnoe and Kinoin}(2000)}]{Curnoe2000}
\bibinfo{author}{\bibfnamefont{S.}~\bibnamefont{Curnoe}} \bibnamefont{and}
  \bibinfo{author}{\bibfnamefont{K.~A.} \bibnamefont{Kinoin}},
  \bibinfo{journal}{Phys. Rev. B} \textbf{\bibinfo{volume}{61}},
  \bibinfo{pages}{15714} (\bibinfo{year}{2000}).

\bibitem[{\citenamefont{Mizumaki et~al.}(2009)\citenamefont{Mizumaki, Tsutsui,
  and Iga}}]{Mizumaki2009}
\bibinfo{author}{\bibfnamefont{M.}~\bibnamefont{Mizumaki}},
  \bibinfo{author}{\bibfnamefont{S.}~\bibnamefont{Tsutsui}}, \bibnamefont{and}
  \bibinfo{author}{\bibfnamefont{F.}~\bibnamefont{Iga}}, \bibinfo{journal}{J.
  Phys.: Conf. Series} \textbf{\bibinfo{volume}{176}}, \bibinfo{pages}{012034}
  (\bibinfo{year}{2009}).

\bibitem[{\citenamefont{Xia et~al.}(2014)\citenamefont{Xia, Jiang, Ye, Wang,
  Zhang, Chen, Niu, Xu, Chen, Chen et~al.}}]{Feng2014}
\bibinfo{author}{\bibfnamefont{M.}~\bibnamefont{Xia}},
  \bibinfo{author}{\bibfnamefont{J.}~\bibnamefont{Jiang}},
  \bibinfo{author}{\bibfnamefont{Z.~R.} \bibnamefont{Ye}},
  \bibinfo{author}{\bibfnamefont{Y.~H.} \bibnamefont{Wang}},
  \bibinfo{author}{\bibfnamefont{Y.}~\bibnamefont{Zhang}},
  \bibinfo{author}{\bibfnamefont{S.~D.} \bibnamefont{Chen}},
  \bibinfo{author}{\bibfnamefont{X.~H.} \bibnamefont{Niu}},
  \bibinfo{author}{\bibfnamefont{D.~F.} \bibnamefont{Xu}},
  \bibinfo{author}{\bibfnamefont{F.}~\bibnamefont{Chen}},
  \bibinfo{author}{\bibfnamefont{X.~H.} \bibnamefont{Chen}},
  \bibnamefont{et~al.}, \bibinfo{journal}{Sci. Rep.}
  \textbf{\bibinfo{volume}{4}}, \bibinfo{pages}{5999} (\bibinfo{year}{2014}).

\bibitem[{\citenamefont{Xu et~al.}()\citenamefont{Xu, Matt, Pomjakushina, Dil,
  Landolt, Ma, Shi, Dhaka, Plumb, Radovic et~al.}}]{Ming2014}
\bibinfo{author}{\bibfnamefont{N.}~\bibnamefont{Xu}},
  \bibinfo{author}{\bibfnamefont{C.~E.} \bibnamefont{Matt}},
  \bibinfo{author}{\bibfnamefont{E.}~\bibnamefont{Pomjakushina}},
  \bibinfo{author}{\bibfnamefont{J.~H.} \bibnamefont{Dil}},
  \bibinfo{author}{\bibfnamefont{G.}~\bibnamefont{Landolt}},
  \bibinfo{author}{\bibfnamefont{J.-Z.} \bibnamefont{Ma}},
  \bibinfo{author}{\bibfnamefont{X.}~\bibnamefont{Shi}},
  \bibinfo{author}{\bibfnamefont{R.~S.} \bibnamefont{Dhaka}},
  \bibinfo{author}{\bibfnamefont{N.~C.} \bibnamefont{Plumb}},
  \bibinfo{author}{\bibfnamefont{M.}~\bibnamefont{Radovic}},
  \bibnamefont{et~al.}, \bibinfo{note}{arXiv:1405.0165 (2014)}.

\bibitem[{\citenamefont{Neupane et~al.}()\citenamefont{Neupane, Xu, Alidoust,
  Bian, Kim, liu, Belopolski, and. H.-T.~Jeng, Durakiewicz, Lin
  et~al.}}]{Hasan2014}
\bibinfo{author}{\bibfnamefont{M.}~\bibnamefont{Neupane}},
  \bibinfo{author}{\bibfnamefont{S.-Y.} \bibnamefont{Xu}},
  \bibinfo{author}{\bibfnamefont{N.}~\bibnamefont{Alidoust}},
  \bibinfo{author}{\bibfnamefont{G.}~\bibnamefont{Bian}},
  \bibinfo{author}{\bibfnamefont{D.~J.} \bibnamefont{Kim}},
  \bibinfo{author}{\bibfnamefont{C.}~\bibnamefont{liu}},
  \bibinfo{author}{\bibfnamefont{I.}~\bibnamefont{Belopolski}},
  \bibinfo{author}{\bibfnamefont{T.-R.~C.} \bibnamefont{and. H.-T.~Jeng}},
  \bibinfo{author}{\bibfnamefont{T.}~\bibnamefont{Durakiewicz}},
  \bibinfo{author}{\bibfnamefont{H.}~\bibnamefont{Lin}}, \bibnamefont{et~al.},
  \bibinfo{note}{arXiv:1404.6814 (2014)}.

\bibitem[{\citenamefont{Speer et~al.}(2006)\citenamefont{Speer, Tang, and
  Chiang}}]{Speer2006}
\bibinfo{author}{\bibfnamefont{N.~J.} \bibnamefont{Speer}},
  \bibinfo{author}{\bibfnamefont{S.-J.} \bibnamefont{Tang}}, \bibnamefont{and}
  \bibinfo{author}{\bibfnamefont{T.-C.} \bibnamefont{Chiang}},
  \bibinfo{journal}{Science} \textbf{\bibinfo{volume}{314}},
  \bibinfo{pages}{804} (\bibinfo{year}{2006}).

\bibitem[{\citenamefont{Santander-Syro
  et~al.}(2011)\citenamefont{Santander-Syro, Copie, Kondo, Fortuna,
  Pailh{\`e}s, Weht, Qiu, Bertran, Nicolaou, Taleb-Ibrahimi
  et~al.}}]{Santander2011}
\bibinfo{author}{\bibfnamefont{A.~F.} \bibnamefont{Santander-Syro}},
  \bibinfo{author}{\bibfnamefont{O.}~\bibnamefont{Copie}},
  \bibinfo{author}{\bibfnamefont{T.}~\bibnamefont{Kondo}},
  \bibinfo{author}{\bibfnamefont{F.}~\bibnamefont{Fortuna}},
  \bibinfo{author}{\bibfnamefont{S.}~\bibnamefont{Pailh{\`e}s}},
  \bibinfo{author}{\bibfnamefont{R.}~\bibnamefont{Weht}},
  \bibinfo{author}{\bibfnamefont{X.~G.} \bibnamefont{Qiu}},
  \bibinfo{author}{\bibfnamefont{F.}~\bibnamefont{Bertran}},
  \bibinfo{author}{\bibfnamefont{A.}~\bibnamefont{Nicolaou}},
  \bibinfo{author}{\bibfnamefont{A.}~\bibnamefont{Taleb-Ibrahimi}},
  \bibnamefont{et~al.}, \bibinfo{journal}{Nature (London)}
  \textbf{\bibinfo{volume}{469}}, \bibinfo{pages}{189} (\bibinfo{year}{2011}).

\bibitem[{\citenamefont{Denlinger
  et~al.}(2002{\natexlab{b}})\citenamefont{Denlinger, Clack, Allen, Gweon,
  Poirier, Olson, , Sarrao, Bianchi, and Fisk}}]{Denlinger2002}
\bibinfo{author}{\bibfnamefont{J.~D.} \bibnamefont{Denlinger}},
  \bibinfo{author}{\bibfnamefont{J.~A.} \bibnamefont{Clack}},
  \bibinfo{author}{\bibfnamefont{J.~W.} \bibnamefont{Allen}},
  \bibinfo{author}{\bibfnamefont{G.-H.} \bibnamefont{Gweon}},
  \bibinfo{author}{\bibfnamefont{D.~M.} \bibnamefont{Poirier}},
  \bibinfo{author}{\bibfnamefont{C.~G.} \bibnamefont{Olson}}, ,
  \bibinfo{author}{\bibfnamefont{J.~L.} \bibnamefont{Sarrao}},
  \bibinfo{author}{\bibfnamefont{A.~D.} \bibnamefont{Bianchi}},
  \bibnamefont{and} \bibinfo{author}{\bibfnamefont{Z.}~\bibnamefont{Fisk}},
  \bibinfo{journal}{Phys. Rev. Lett.} \textbf{\bibinfo{volume}{89}},
  \bibinfo{pages}{157601} (\bibinfo{year}{2002}{\natexlab{b}}).

\bibitem[{\citenamefont{Meevasana et~al.}(2011)\citenamefont{Meevasana, King,
  He, Mo, Hashimoto, Tamai, Songsiriritthigul, Baumberger, and
  Shen}}]{Meevasana2011}
\bibinfo{author}{\bibfnamefont{W.}~\bibnamefont{Meevasana}},
  \bibinfo{author}{\bibfnamefont{P.~D.~C.} \bibnamefont{King}},
  \bibinfo{author}{\bibfnamefont{R.~H.} \bibnamefont{He}},
  \bibinfo{author}{\bibfnamefont{S.-K.} \bibnamefont{Mo}},
  \bibinfo{author}{\bibfnamefont{M.}~\bibnamefont{Hashimoto}},
  \bibinfo{author}{\bibfnamefont{A.}~\bibnamefont{Tamai}},
  \bibinfo{author}{\bibfnamefont{P.}~\bibnamefont{Songsiriritthigul}},
  \bibinfo{author}{\bibfnamefont{F.}~\bibnamefont{Baumberger}},
  \bibnamefont{and} \bibinfo{author}{\bibfnamefont{Z.-X.} \bibnamefont{Shen}},
  \bibinfo{journal}{Nature Mater.} \textbf{\bibinfo{volume}{10}},
  \bibinfo{pages}{114} (\bibinfo{year}{2011}).

\bibitem[{\citenamefont{King et~al.}(2011)\citenamefont{King, Hatch, Bianchi,
  Ovsyannikov, Lupulescu, Guan, Mi, Rienks, Fink, Lindblad et~al.}}]{King2011}
\bibinfo{author}{\bibfnamefont{P.~D.~C.} \bibnamefont{King}},
  \bibinfo{author}{\bibfnamefont{R.~C.} \bibnamefont{Hatch}},
  \bibinfo{author}{\bibfnamefont{M.}~\bibnamefont{Bianchi}},
  \bibinfo{author}{\bibfnamefont{R.}~\bibnamefont{Ovsyannikov}},
  \bibinfo{author}{\bibfnamefont{C.}~\bibnamefont{Lupulescu}},
  \bibinfo{author}{\bibfnamefont{D.}~\bibnamefont{Guan}},
  \bibinfo{author}{\bibfnamefont{J.~L.} \bibnamefont{Mi}},
  \bibinfo{author}{\bibfnamefont{E.~D.~L.} \bibnamefont{Rienks}},
  \bibinfo{author}{\bibfnamefont{J.}~\bibnamefont{Fink}},
  \bibinfo{author}{\bibfnamefont{A.}~\bibnamefont{Lindblad}},
  \bibnamefont{et~al.}, \bibinfo{journal}{Phys. Rev. Lett.}
  \textbf{\bibinfo{volume}{107}}, \bibinfo{pages}{096802}
  (\bibinfo{year}{2011}).

\bibitem[{\citenamefont{Harada et~al.}(1999)\citenamefont{Harada, Sekiyama,
  Suga, Imada, Muro, Jung, Matsuda, Takagi, Ochiai, Suzuki
  et~al.}}]{Harada1999}
\bibinfo{author}{\bibfnamefont{H.}~\bibnamefont{Harada}},
  \bibinfo{author}{\bibfnamefont{A.}~\bibnamefont{Sekiyama}},
  \bibinfo{author}{\bibfnamefont{S.}~\bibnamefont{Suga}},
  \bibinfo{author}{\bibfnamefont{S.}~\bibnamefont{Imada}},
  \bibinfo{author}{\bibfnamefont{T.}~\bibnamefont{Muro}},
  \bibinfo{author}{\bibfnamefont{R.-J.} \bibnamefont{Jung}},
  \bibinfo{author}{\bibfnamefont{K.}~\bibnamefont{Matsuda}},
  \bibinfo{author}{\bibfnamefont{H.}~\bibnamefont{Takagi}},
  \bibinfo{author}{\bibfnamefont{A.}~\bibnamefont{Ochiai}},
  \bibinfo{author}{\bibfnamefont{T.}~\bibnamefont{Suzuki}},
  \bibnamefont{et~al.}, \bibinfo{journal}{J. Phys. Soc. Japan}
  \textbf{\bibinfo{volume}{68}}, \bibinfo{pages}{2844} (\bibinfo{year}{1999}).

\bibitem[{\citenamefont{Joyce et~al.}(1996)\citenamefont{Joyce, Andrews, Arko,
  Bartlett, Blythe, Olson, Benning, Canfield, and Poirier}}]{Joyce1996}
\bibinfo{author}{\bibfnamefont{J.~J.} \bibnamefont{Joyce}},
  \bibinfo{author}{\bibfnamefont{A.~B.} \bibnamefont{Andrews}},
  \bibinfo{author}{\bibfnamefont{A.~J.} \bibnamefont{Arko}},
  \bibinfo{author}{\bibfnamefont{R.~J.} \bibnamefont{Bartlett}},
  \bibinfo{author}{\bibfnamefont{R.~I.~R.} \bibnamefont{Blythe}},
  \bibinfo{author}{\bibfnamefont{C.~G.} \bibnamefont{Olson}},
  \bibinfo{author}{\bibfnamefont{P.~J.} \bibnamefont{Benning}},
  \bibinfo{author}{\bibfnamefont{P.~C.} \bibnamefont{Canfield}},
  \bibnamefont{and} \bibinfo{author}{\bibfnamefont{D.~M.}
  \bibnamefont{Poirier}}, \bibinfo{journal}{Phys. Rev. B}
  \textbf{\bibinfo{volume}{54}}, \bibinfo{pages}{17515} (\bibinfo{year}{1996}).

\bibitem[{\citenamefont{Schmitt et~al.}(2001)\citenamefont{Schmitt, St\"{u}ckl,
  Ripplinger, and Albert}}]{Schmitt2001}
\bibinfo{author}{\bibfnamefont{K.}~\bibnamefont{Schmitt}},
  \bibinfo{author}{\bibfnamefont{C.}~\bibnamefont{St\"{u}ckl}},
  \bibinfo{author}{\bibfnamefont{H.}~\bibnamefont{Ripplinger}},
  \bibnamefont{and} \bibinfo{author}{\bibfnamefont{B.}~\bibnamefont{Albert}},
  \bibinfo{journal}{Solid State Sci.} \textbf{\bibinfo{volume}{3}},
  \bibinfo{pages}{321} (\bibinfo{year}{2001}).

\bibitem[{\citenamefont{Mackinnon et~al.}(2013)\citenamefont{Mackinnon, Alarco,
  and Talbot}}]{Mackinnon2013}
\bibinfo{author}{\bibfnamefont{I.~D.~R.} \bibnamefont{Mackinnon}},
  \bibinfo{author}{\bibfnamefont{J.~A.} \bibnamefont{Alarco}},
  \bibnamefont{and} \bibinfo{author}{\bibfnamefont{P.~C.}
  \bibnamefont{Talbot}}, \bibinfo{journal}{Modeling and Numer. Simul. of Mater.
  Sci.} \textbf{\bibinfo{volume}{3}}, \bibinfo{pages}{158}
  (\bibinfo{year}{2013}).

\bibitem[{\citenamefont{Alarco et~al.}(2014)\citenamefont{Alarco, Talbot, and
  Mackinnon}}]{Alarco2014}
\bibinfo{author}{\bibfnamefont{J.~A.} \bibnamefont{Alarco}},
  \bibinfo{author}{\bibfnamefont{P.~C.} \bibnamefont{Talbot}},
  \bibnamefont{and} \bibinfo{author}{\bibfnamefont{I.~D.~R.}
  \bibnamefont{Mackinnon}}, \bibinfo{journal}{Modeling and Numer. Simul. of
  Mater. Sci.} \textbf{\bibinfo{volume}{4}}, \bibinfo{pages}{53}
  (\bibinfo{year}{2014}).

\bibitem[{\citenamefont{Bao et~al.}(2013)\citenamefont{Bao, Tegus, Zhang,
  Zhang, and Huang}}]{Bao2013}
\bibinfo{author}{\bibfnamefont{L.~H.} \bibnamefont{Bao}},
  \bibinfo{author}{\bibfnamefont{O.}~\bibnamefont{Tegus}},
  \bibinfo{author}{\bibfnamefont{J.~X.} \bibnamefont{Zhang}},
  \bibinfo{author}{\bibfnamefont{X.}~\bibnamefont{Zhang}}, \bibnamefont{and}
  \bibinfo{author}{\bibfnamefont{Y.~K.} \bibnamefont{Huang}},
  \bibinfo{journal}{J. Alloys and Compounds} \textbf{\bibinfo{volume}{558}},
  \bibinfo{pages}{39} (\bibinfo{year}{2013}).

\end{thebibliography}
\end{document}